\documentclass{sig-alternate}
\usepackage{times}
\usepackage{helvet}
\usepackage{courier}

\usepackage[draft]{hyperref}

\usepackage[utf8]{inputenc}
\usepackage{url}
\usepackage{microtype}
\usepackage{graphicx}
\usepackage{booktabs}
\usepackage{tabulary}
\usepackage{subcaption}
\usepackage{balance}

\usepackage[backend=bibtex,style=numeric,hyperref=true,natbib]{biblatex}
\bibliography{wiki_gender_bias}

\usepackage{enumitem}
\setlist{nolistsep}

\usepackage[bottom]{footmisc}

\usepackage{algorithmicx}
\usepackage{algorithm}
\usepackage{algpseudocode}

\renewcommand{\cite}{\citep}

\newcommand{\ie}{\emph{i.\,e.}}

\newcommand{\eg}{\emph{e.\,g.}}
 
\newcommand{\wrt}{with respect to }

\newfont{\mycrnotice}{ptmr8t at 7pt}
\newfont{\myconfname}{ptmri8t at 7pt}
\let\crnotice\mycrnotice%
\let\confname\myconfname%

\usepackage{ifthen}
\newboolean{inthesis}
\setboolean{inthesis}{false}

{\ifthenelse{ \boolean{inthesis} }{}{}
{\ifthenelse{ \boolean{inthesis} }{\newcommand{\bigtable}{sidewaystable}}{\newcommand{\bigtable}{table*}}
{\ifthenelse{ \boolean{inthesis} }{\newcommand{\widefigure}{figure}}{\newcommand{\widefigure}{figure*}}
{\ifthenelse{ \boolean{inthesis} }{}{}
{\ifthenelse{ \boolean{inthesis} }{\newcommand{\spara}[1]{\subsection{#1}}}{\newcommand{\spara}[1]{\smallskip\noindent{\bf #1.}}}

\newcommand{\superscript}[1]{\ensuremath{^{\textrm{#1}}}}
\def\sharedaffiliation{\end{tabular}\newline\begin{tabular}{c}}

\numberofauthors{3}
\author{
\alignauthor Eduardo Graells-Garrido\superscript{1,2} \\
\alignauthor Mounia Lalmas\superscript{3} \\
\alignauthor Filippo Menczer\superscript{4,5} \\
\sharedaffiliation
\begin{tabular}{ccccccccc}
\affaddr{{\superscript{1}}Web Research Group{\ }} & & \affaddr{{\superscript{2}}Telefónica I+D{\ }} & & \affaddr{{\superscript{3}}Yahoo Labs{\ }}     & & \affaddr{{\superscript{4}}Yahoo Labs{\ }} & & \affaddr{{\superscript{5}}Indiana University{\ }}\\
\affaddr{Universitat Pompeu Fabra}    & & \affaddr{Santiago, Chile}         & & \affaddr{London, UK}              & & \affaddr{Sunnyvale, USA} & & \affaddr{Bloomington, USA}\\
\affaddr{Barcelona, Spain}            & &                                   & &                                   & &  & &  \\
\end{tabular}
}

\title{First Women, Second Sex: Gender Bias in Wikipedia}

\clubpenalty=10000
\widowpenalty=10000

\toappear{This is the author's version of this paper. To be presented at the ACM Conference on Hypertext and Social Media 2015.}

\begin{document}
 
\maketitle

\begin{abstract}
Contributing to history has never been as easy as it is today. Anyone with access to the Web is able to play a part on Wikipedia, an open and free encyclopedia. Wikipedia, available in many languages, is one of the most visited websites in the world and arguably one of the primary sources of knowledge on the Web. 
However, not \textit{everyone} is contributing to Wikipedia from a diversity point of view; several groups are severely underrepresented. 
One of those groups is \textit{women}, who make up approximately 16\% of the current contributor community, meaning that most of the content is written by men.
In addition, although there are specific guidelines of verifiability, notability, and neutral point of view that must be adhered by Wikipedia content, these guidelines are supervised and enforced by men.

In this {\ifthenelse{\boolean{inthesis}}{chapter}{paper}}, we propose that gender bias is not about participation and representation only, but also about characterization of women.
We approach the analysis of gender bias by defining a methodology for comparing the characterizations of men and women in biographies in three aspects: meta-data, language, and network structure.
Our results show that, indeed, there are differences in characterization and structure. 
Some of these differences are reflected from the off-line world documented by Wikipedia, but other differences can be attributed to gender bias in Wikipedia content. 
We contextualize these differences in feminist theory and discuss their implications for Wikipedia policy.
\end{abstract}

\category{H.3.4}{Information Storage and Retrieval}{Systems and Software}[Information networks]

\keywords{Wikipedia; Gender; Gender Bias; Computational Linguistics.}

\section{Introduction}
%

Today's Web creates opportunities for global and democratic media, where everyone has a voice. One of the most visible examples is Wikipedia, an open encyclopedia where anyone can contribute content. 
In contrast to traditional encyclopedias, where a staff of experts in specific areas takes care of writing, editing and validating content, in Wikipedia these tasks are performed by a community of volunteers.
Whether or not this \textit{open source} approach provides reliable and accurate content \cite{giles2005internet,rosenzweig2006can}, Wikipedia has gained unprecedented reach. Indeed, Wikipedia was the 7th most visited website during 2014.\footnote{\url{http://www.alexa.com/siteinfo/wikipedia.org}}
An extensive body of research builds upon Wikipedia \cite{okoli2014wikipedia}, covering topics like participation, structured data, and analysis of historical figures, among others.

In theory, by following its guidelines about verifiability, notability, and neutral point of view, Wikipedia should be an unbiased source of knowledge. In practice, the community of Wikipedians is not diverse, and contributors are inherently biased.
One group that is severely underrepresented in Wikipedia is women, who represent only 16\% of editors \citep{hill2013wikipedia}. 
This disparity has been called the \textit{gender gap} in Wikipedia, and has been studied from several perspectives to understand why more women do not join Wikipedia, and what can be done about it. 
It is a problem because reportedly women are not being treated as equals to men in the community \cite{lam2011wp}, and potentially, in content.
For instance, \citet{filipacchi2013wikipedia} described a controversy where women novelists started to be excluded from the category \textit{``American Novelists''} to be included in the specific category \textit{``American Women Novelists.''}

Instead of focusing on the participatory \textit{gender gap}, we focus on how women are characterized in Wikipedia articles, to assess whether gender bias from the off-line world extends to Wikipedia content, and to identify biases exhibited by Wikipedians in the characterization of women and of their historical significance.
The research questions that drive our work are:

\begin{quote}
 \textit{Is there a gender bias in user-generated characterizations of men and women in Wikipedia? \\
 If so, how to identify and quantify it? How to explain it based on social theory?}
\end{quote}

The study of biases in Wikipedia is not new. \citet{hecht2009measuring} defined the notion of \textit{self-focus bias} to study the cultural biases present in Wikipedia from a ``hyperlingual'' approach.
Having the gender gap in mind, we focus on gender bias not only to quantify it, but to understand what could be causing it.
As a first approach to the problem, we focus on the English language to be able to analyze our results in terms of western feminist theories from the social sciences.

In the book \textit{The Second Sex}, Simone de Beauvoir widely discusses different aspects of women oppression and their historical significance. She wrote in 1949:  
\textit{``it is not women's inferiority that has determined their historical insignificance: it is their historical insignificance that has doomed them to inferiority''} \cite{de2012second}.
More than 60 years later, almost anyone with access to the Web can contribute to the writing of history, thanks to Wikipedia.
The scale of Wikipedia, as well as its openness, allows us to perform a quantitative analysis of how women are characterized in Wikipedia in comparison to men.
Encyclopedias characterize men and women in many ways, \eg, in terms of their lives and the events in which they participated or were relevant. 
We concentrate on \emph{biographies} because they are a good source to study gender bias, given that each article is about a specific person.
We propose three dimensions along which to perform our analysis: \textit{meta-data}, \textit{language}, and \textit{network structure}. 
This leads to three major findings:

\begin{enumerate}
\item Differences in meta-data are coherent with results in previous work, where women biographies were found to contain more marriage-related events than men's. 

\item Sex-related content is more frequent in women biographies than men's, while cognition-related content is more highlighted in men biographies than women's.

\item A strong bias in the linking patterns results in a network structure in which articles about men are disproportionately more central than articles about women. 
\end{enumerate}

The main contributions of this work are methods to quantify gender bias in user generated content, a contextualization of differences found in terms of feminist theory, and a discussion of the implications of our findings for informing policy design in Wikipedia. 
As said earlier, we focus on the English Wikipedia, but our methods are generalizable to other languages and platforms.

\section{Background}
Research on the community structure and evolution of Wikipedia has been prominent.
In its first steps, the focus was on growth \cite{almeida2007evolution} and dynamics \cite{ratkiewicz2010characterizing}, without attention toward gender.
Later, it was found that there is a gender gap, as Wikipedia has fewer contributions from women, and women stop contributing earlier than men \cite{lam2011wp}.
There are differences in how genders behave. 
For instance, men and women communicate differently in the inner communication channels in Wikipedia \cite{laniado2012emotions}: they focus on different topics \cite{lam2011wp} and the level of content revision differs by gender but also by amount of activity \cite{antin2011gender}.
In addition, \citet{lam2011wp} found that women are more \textit{reverted} than men (\textit{i.e.}, their contributions are discarded), and reportedly women contribute less because of aggressive behavior toward them \cite{collier2012conflict,wikisurvey}.
Efforts have been made to build a more welcoming community and to encourage participation \cite{morgan2013tea,ciampaglia2014moodbar}, and Wikimedia itself encourages initiatives like \textit{WikiWomen's Collaborative}.\footnote{\url{http://meta.wikimedia.org/wiki/WikiWomen's_Collaborative}} 

Content-wise, the study of biographies in Wikipedia enables cultural comparisons of coverage \cite{callahan2011cultural}, as well as the construction of social networks of historical (and current) figures \cite{aragon2012biographical}. 
Although bias in content has been addressed before through \textit{self-focus bias} \cite{hecht2009measuring}, such bias has been measured at large-scale only in terms of culture, not gender.
\citet{lam2011wp} found that coverage of ``female topics'' was inferior to ``male topics '' when classifying topics as ``male'' or ``female'' according to the people who contributed to them. 
\citet{reagle2011gender} found that in characterization of women, in comparison to  commercial encyclopedias like \textit{Britannica}, Wikipedia has better coverage of notable profiles, although this coverage is quite low and it is still biased towards men.
\citet{bamman2014unsupervised} found that women biographies are more likely to include marriage or divorce events.

Addressing the gender gap from a content perspective may help to improve the quality and value of the content. 
Currently, focus on quality in Wikipedia has been about predicting article quality \cite{anderka2012predicting,flekova2014makes}.
However, focusing on quality without considering readers does not give the whole picture, as Wikipedia readers are not necessarily interested in the same topics as contributors \cite{lehmann2014reader} and might have a different concept of quality.
Moreover, in our context, \citet{flekova2014makes} found that quality of biographies is assessed differently depending on the gender of the portrayed person.
Is it because the raters were biased? Or is it because biographies were written differently?
Our hypothesis is that biographies are written differently, an idea inspired by seminal work about how women are characterized by language \cite{lakoff1972language}.

To study differences in text, word frequency is commonly used.
Word frequency follows Zipf's law \cite{zipf1949human,serrano2009modeling}{\ifthenelse{\boolean{inthesis}}{, 
an empirical distribution found in many languages \cite{piantadosi2014zipf}}{}}. 
An interesting property of Zipf distributions in language is that small sets of words that are semantically or categorically related also follow a Zipf distribution \cite{piantadosi2014zipf}.
This property implies that, given two subsets of words that are related semantically or categorically, their frequency distributions can be compared.
Thus, we compare frequency distributions according to gender for several semantic categories derived from the \textit{Linguistic Inquiry and Word Count} (LIWC) dictionary. LIWC studies \textit{``emotional, cognitive, structural, and process components present in individuals' verbal and written speech samples''} \cite{pennebaker2001linguistic}.
It has been used to analyze interactions between Wikipedia contributors \cite{iosub2014emotions} and article content \wrt emotions \cite{ferron2012psychological}.
In a context similar to ours, \citet{schmader2007linguistic} used LIWC to quantify differences in characterization of women and men in recommendation letters.

In our work, we quantify gender bias in Wikipedia's characterization of men and women through their biographies.
To do so we approach three different dimensions of biographies, which we analyze in different sections on this {\ifthenelse{\boolean{inthesis}}{chapter}{paper}}: \textit{meta-data}, provided by the structured version of Wikipedia, DBPedia \cite{dbpediaswj};
\textit{language}, considering how frequent are words and concepts \cite{serrano2009modeling};
and \textit{network structure}.
In terms of network structure, we build a biography network \cite{aragon2012biographical} in which we estimate PageRank, a measure of node centrality based on network connectivity \cite{brin1998anatomy,Fortunato08internetmath}. 
In similar contexts, PageRank has been used to provide an approximation of historical importance \cite{aragon2012biographical,whosbigger} and to study the bias leading to the gender gap \cite{whosbigger}.
We measure bias in link formation by comparing the importance given by PageRank in the biography network with those of null models, \ie, graphs that are unbiased by construction but that maintain certain properties of the source biography network. 

\section{Dataset}
To study gender bias in Wikipedia, we consider three freely available data sources:

\begin{enumerate}
 \item The DBPedia 2014 dataset \cite{dbpediaswj}.\footnote{\url{http://wiki.dbpedia.org/Downloads2014}}
 \item The Wikipedia English Dump of October 2014.\footnote{\url{https://dumps.wikimedia.org/enwiki/20141008/}}
 \item Inferred gender for Wikipedia biographies by \citet{bamman2014unsupervised}.\footnote{\url{http://www.ark.cs.cmu.edu/bio/}}
\end{enumerate}

\begin{figure}[tb]
\centering
\includegraphics[width=0.4\linewidth]{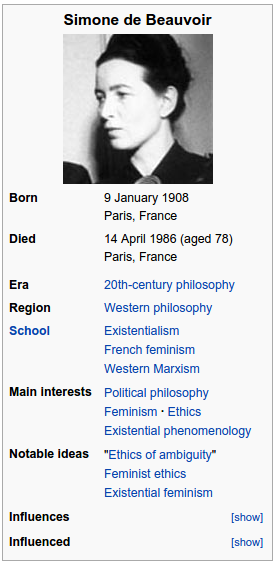}
\caption{\textit{Infobox} from the biography article of Simone de Beauvoir.}
\label{fig:infobox}
\end{figure}

\spara{DBPedia and Meta-data}
DBPedia is a structured version of Wikipedia that provides meta-data for articles, normalized article URIs (\textit{Uniform Resource Identifiers}), and normalized links between articles (taking care of redirections). It provides a shallow hierarchy of classes, which includes a \textit{Person} category. 
To provide the structured meta-data, DBPedia processes from the content of infoboxes in Wikipedia articles. 
Infoboxes are template-based specifications for specific kinds of articles.
When DBPedia detects an infobox with a template that matches those of a person, it assigns the article to the \textit{Person} class, and to a specific subclass if applicable (\eg, \textit{Artist}).
{\ifthenelse{\boolean{inthesis}}{%
For instance, Figure \ref{fig:infobox} displays the infobox of \textit{Simone de Beauvoir}\footnote{\url{https://en.wikipedia.org/wiki/Simone_de_Beauvoir}} \cite{de2012second}. 
}{%
For instance, Figure \ref{fig:infobox} displays the infobox of Simone de Beauvoir. 
}}%
The infobox contains specific meta-data pertinent to a biography, such as date and place of birth, but it does not include gender (in specific cases it does, see ``Inferred Gender'' next).
DBPedia maps infobox properties to specific fields in a person's meta-data. 
These properties are not always available in the infobox templates, and do not always have a standardized name. DBPedia, whenever possible, normalizes both attribute keys and attribute values.

\spara{Wikipedia Biographies}
We consider two versions of the biographies: the overview and the full text.
We analyze both in different contexts: in the overview we analyze the full vocabulary employed, while in the full text we analyze only the words pertaining to the LIWC dictionaries.
The overview is described by Wikipedia as \textit{``an introduction to the article and a summary of its most important aspects. It should be able to stand alone as a concise overview.'' }
Since those aspects are subjective, the introduction content is a good proxy for any potential biases expressed by Wikipedia contributors. 
At the same time we avoid potential noise included in the full biography text from elements like quotations and the filmography of a given actor/actress. 
In both cases (overview and full content), template markup is removed from analysis.

\spara{Inferred Gender}
To obtain gender meta-data for biographies, we match article URIs with the dataset by \citet{bamman2014unsupervised}, which contains inferred gender for biographies based on the number of grammatically gendered words (\ie, \textit{he}, \textit{she}, \textit{him}, \textit{her}, etc.) present in the article text. 
\citet{bamman2014unsupervised} tested their method in a random set of 500 biographies, providing 100\% precision and 97.6\% recall. 
This method has also been used before by \citet{reagle2011gender} and DBPedia itself \cite{dbpediaswj}, making DBPedia to include gender meta-data in some cases.
However, note that the genders considered in these datasets (and thus, in this work) are only \textit{male} and \textit{female}.

\section{Meta-Data Properties}
In our first analysis we estimate the proportion of women in Wikipedia.
We analyze meta-data by comparing how men and women proportionally have several attributes in the data from DBPedia.

\spara{Presence and Proportion According to Class}
DBPedia estimates the length (in characters) and provides the connectivity of articles.
Of the set of 1,445,021 biographies (articles in the DBPedia Person class), 893,380 (61.82\%) have gender meta-data. 
Of those, only 15.5\% are about women.

The mean article length is 5,955 characters for men and 6,013 characters for women (a significant difference according to a t-test for independent samples: 
$p < 0.01$, Cohen's $d = 0.01$). 
The mean out-degrees (number of links) of 42.1 for men and 39.4 for women also differ significantly ($p < 0.001$, Cohen's $d = 0.06$).
{\ifthenelse{\boolean{inthesis}}{%
Table \ref{table:n-biographies-all} displays the number of biographies in the \textit{Person} class, as well as its most common subclasses with their corresponding out-degrees per gender.  
From the table, in comparison to the global proportion of women, the following categories over-represent women: \textit{Artist}, \textit{Royalty}, \textit{FictionalCharacter}, \textit{Noble}, \textit{BeautyQueen}, and \textit{Model}. The others over-represent men.
The differences in length and degree do not hold for all classes, hinting that a study according to semantic categories of people is needed. 
However, in this chapter we focus on the global differences in \textit{Person}.
}{%
Table \ref{table:n-biographies-all} displays the number of biographies in the \textit{Person} class, as well as its most common subclasses.  
Of all biographies in the dataset, 61.82\% have gender meta-data. 
From the table, in comparison to the global proportion of women, the following categories over-represent women: \textit{Artist}, \textit{Royalty}, \textit{FictionalCharacter}, \textit{Noble}, \textit{BeautyQueen}, and \textit{Model}. The others over-represent men.
The differences in length and degree do not hold for all classes, hinting that a study according to semantic categories of people is needed. 
However, in this paper we focus on the global differences in \textit{Person}.
}}%

\begin{table}[tb]
 \scriptsize
 \caption{Number of biographies in the dataset for the Person class and its most common child classes (in terms of biographies with gender). In this and the following tables, we use this legend for $p$-values: \protect\linebreak *** $p < 0.001$, ** $p < 0.01$, * $p < 0.05$. }
 \centering
{\ifthenelse{\boolean{inthesis}}{%
 \input{table_n_biographies_all}
}{%
 \begin{tabulary}{\linewidth}{lRRLL}
\toprule
           Ontology & With gender &  \% Women & OutD. $t$ &  Len. $t$ \\
\midrule
             Person &      893380 &    15.53 &  20.77*** &   -2.65** \\
            Athlete &      187828 &     8.94 &  10.64*** &   -2.83** \\
             Artist &       79690 &    25.14 &  12.95*** &     -0.33 \\
       OfficeHolder &       38111 &    13.04 &  10.97*** &   3.77*** \\
         Politician &       32398 &     8.75 &      1.29 &  -4.02*** \\
     MilitaryPerson &       22769 &     1.67 &      4*** &      1.03 \\
          Scientist &       15853 &     8.79 &   4.91*** &     -0.01 \\
      SportsManager &       11255 &     0.62 &      0.79 &   -2.79** \\
             Cleric &        8949 &     6.34 &    3.23** &      0.02 \\
            Royalty &        7054 &    35.24 &      0.55 &      1.75 \\
              Coach &        5720 &     2.40 &      0.27 &   -2.65** \\
 FictionalCharacter &        4023 &    26.08 &    3.03** &      0.39 \\
              Noble &        3696 &    23.16 &    3.16** &     2.05* \\
           Criminal &        1976 &    12.45 &      1.08 &     -1.69 \\
              Judge &        1949 &    14.88 &   3.93*** &    2.97** \\
\bottomrule
\end{tabulary}

}}
\label{table:n-biographies-all}
\end{table}

\begin{figure}[tb]
\centering
\includegraphics[width=\linewidth]{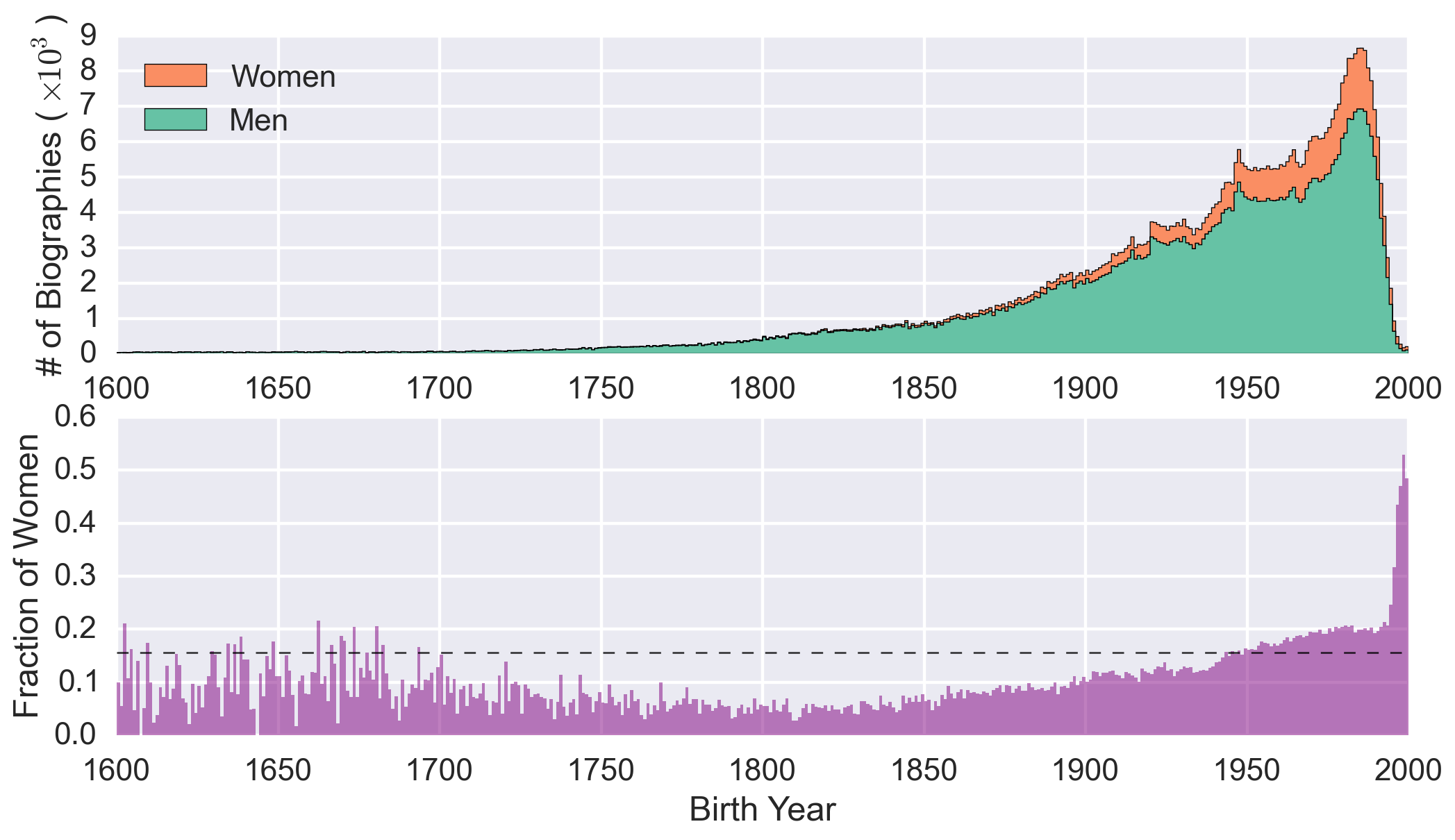}
\caption{Distribution of biographies according to birth year.}
\label{fig:biographies-histogram}
\end{figure}

\begin{figure}[tb]
\centering
\includegraphics[width=\linewidth]{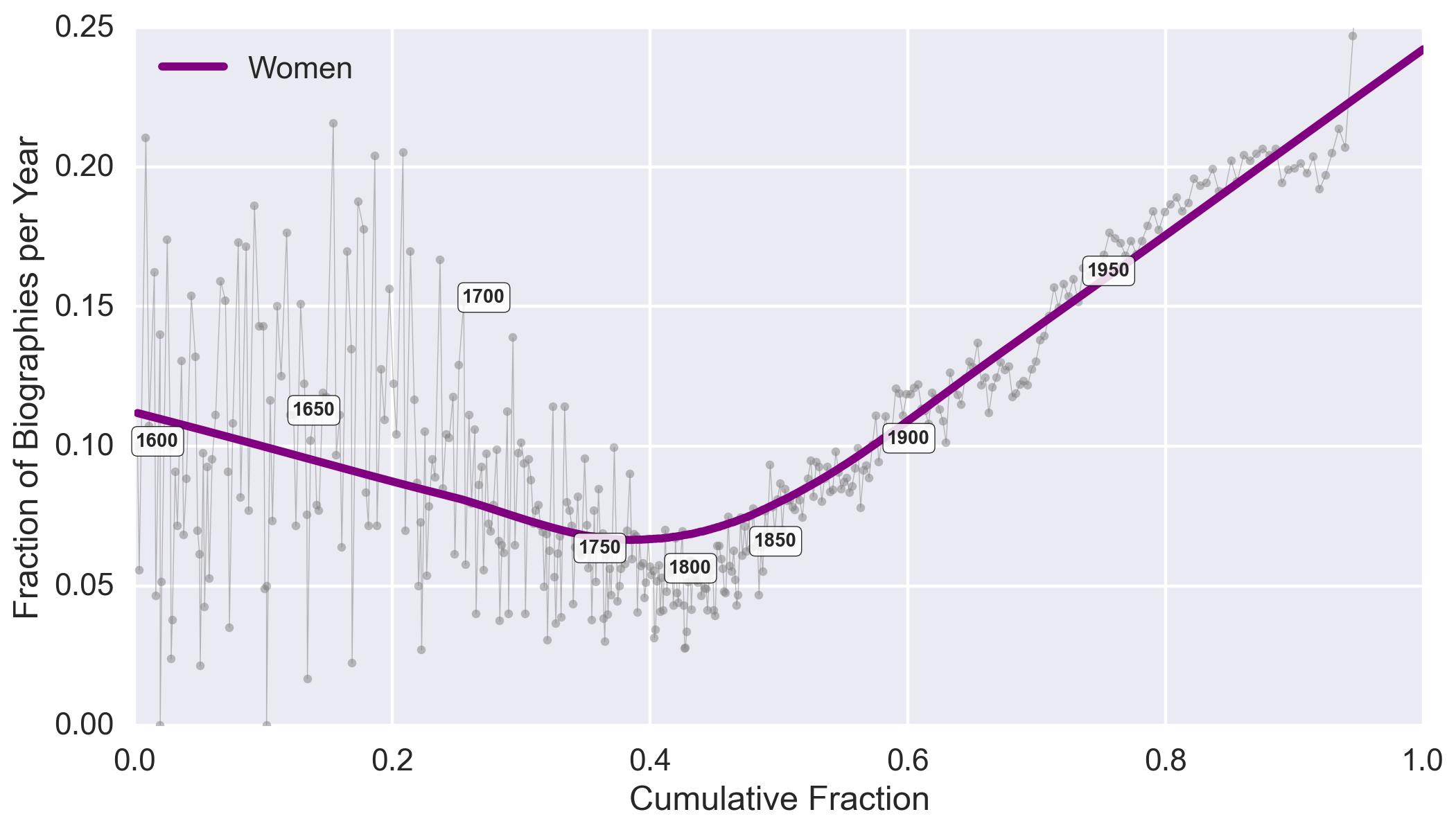}
\caption{Relation between the cumulative fraction of women and the fraction of women per year (dots). The y-axis was truncated to 0.25 for clarity.}
\label{fig:biographies-cumsum}
\end{figure}

\spara{Distribution According to Date of Birth}
Figure \ref{fig:biographies-histogram} displays the distribution of biographies according to their corresponding \textit{birthYear} property, considering only those biographies between years 1600 and 2000 (inclusive). This accounts for 65.48\% of biographies with gender (note that 34.07\% does not have date of birth in meta-data).
The distribution per gender (top chart) shows that most of the biographies of both genders are about people from modern times.
The distribution of fraction of women per year (bottom chart) shows that since the year 1943 the fraction of women is consistently above the global value of 0.155. 
Note that, of the biographies that have date of birth in their meta-data, 53\% are from 1943 until 2000.
To explore the evolution of growth of women presence, in Figure \ref{fig:biographies-cumsum} we display the relationship between the cumulative fraction of biographies and the yearly fraction of biographies of women. 
The chart includes a \textit{LOWESS}\footnote{Locally weighted scatterplot smoothing.} fit of the data, to be able to see the tendency of changes in representation.
This tendency became positive in the period 1750--1800.
These results are discussed in terms of historical significance in the discussion section.

{\ifthenelse{\boolean{inthesis}}{\afterpage{\clearpage}}{}

\spara{Infobox Attributes}
Given that there are different classes of infoboxes, there are many different meta-data attributes than can be included in biographies. 
In total, we identified 340 attributes. 
For each one of them, we counted the number of biographies that contained it, and then compared the relative proportions between genders with a chi-square test.
Only 3.53\% presented statistically significant differences.
Those attributes are displayed in Table \ref{table:biography-meta-data}. 
All of them have large effect sizes (Cohen's $w > 0.5$). 
Inspection allows us to make several observations:

\begin{itemize}
\item Attributes \textit{careerStation}, \textit{formerTeam}, \textit{numberOfMatches}, \textit{position}, \textit{team}, and \textit{years} are more frequent in men.
All these attributes are related to sports, and thus, these differences can be explained by of the prominence of men in sports-related classes (\eg, \textit{Athlete}, \textit{SportsManager} and \textit{Coach} in Table \ref{table:n-biographies-all}).

\item Attributes \textit{deathDate}, \textit{deathYear} are more frequent in men. According to Figure \ref{fig:biographies-histogram}, most women are from recent times, and thus they are presumably still alive.

\item Attribute \textit{birthName} is more frequent in women. Its values refer mostly to the original name of artists, and women have considerable presence in this class (see Table \ref{table:n-biographies-all}). 
In addition, other possible explanation is that, in the case of married women, they usually change their surnames to those of their husbands.

\item Attributes \textit{occupation} and \textit{title} are more frequent in women. \textit{Title} is a description of a person's occupation (the most common are \textit{Actor} and \textit{Actress}), while \textit{occupation} is a DBPedia resource URI (\textit{e.g.}, \url{http://dbpedia.org/resource/Actor}). The infoboxes of sport-related biographies do not contain these attributes because their templates are already indicators of their occupations, and thus, athletes (which are mostly men) do not contain such attributes.
\end{itemize}

The case of the \textit{spouse} attribute is different. The inspection does not offer a direct explanation other than the tendency to include this attribute more in women biographies than in men's. 
For instance, the most common class with the spouse attribute is \textit{Person}, the reference class, with 45\% of the instances of the attribute. 

\begin{table}[tb]
 \scriptsize
 \caption{Proportion of men and women who have the specified attributes in their infoboxes. Proportions were tested with a chi-square test, with effect size estimated using Cohen's $w$.}
\centering
 \begin{tabulary}{\linewidth}{lRRLR}
\toprule
{} &  \% Men &  \% Women &    $\chi^2$ &    w \\
\midrule
birthName       &          4.01 &           11.46 &    4.84* & 0.81 \\
careerStation   &          8.95 &            1.13 &   6.84** & 0.94 \\
deathDate       &         32.82 &           19.35 &    5.53* & 0.64 \\
deathYear       &         44.68 &           25.45 &   8.28** & 0.66 \\
formerTeam      &          4.40 &            0.24 &    3.94* & 0.97 \\
numberOfMatches &          8.60 &            1.06 &    6.61* & 0.94 \\
occupation      &         12.52 &           23.28 &    4.97* & 0.68 \\
position        &         13.62 &            1.68 &  10.46** & 0.94 \\
spouse          &          1.56 &            6.86 &    4.10* & 0.88 \\
team            &         14.06 &            1.97 &  10.39** & 0.93 \\
title           &          9.17 &           19.65 &    5.59* & 0.73 \\
years           &          8.95 &            1.12 &   6.84** & 0.94 \\
\bottomrule
\end{tabulary}

\label{table:biography-meta-data}
\end{table}


\section{Language Properties}

{\ifthenelse{\boolean{inthesis}}{%
\begin{figure}[tbp]
\centering
\includegraphics[width=\linewidth]{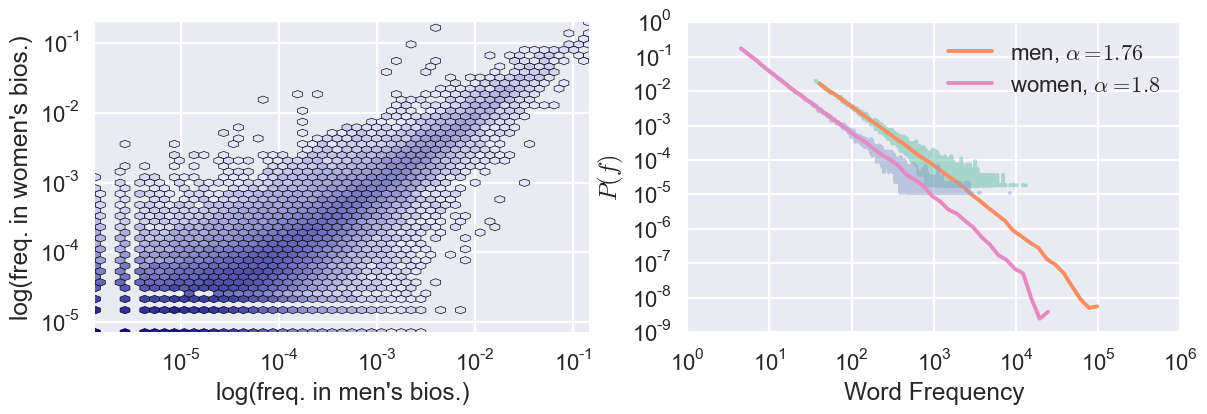}
\caption{A density hexbin plot of word frequencies in men/women's biographies (left), and the PDF of word frequency distribution according to gender (right). 
Fitting to Zipfian distributions with the \textit{powerlaw} library \cite{alstott2014powerlaw} yields the shown exponents.}
\label{fig:word-frequencies}
\end{figure}
}{
\begin{figure}[tb]
\centering
\includegraphics[width=\linewidth]{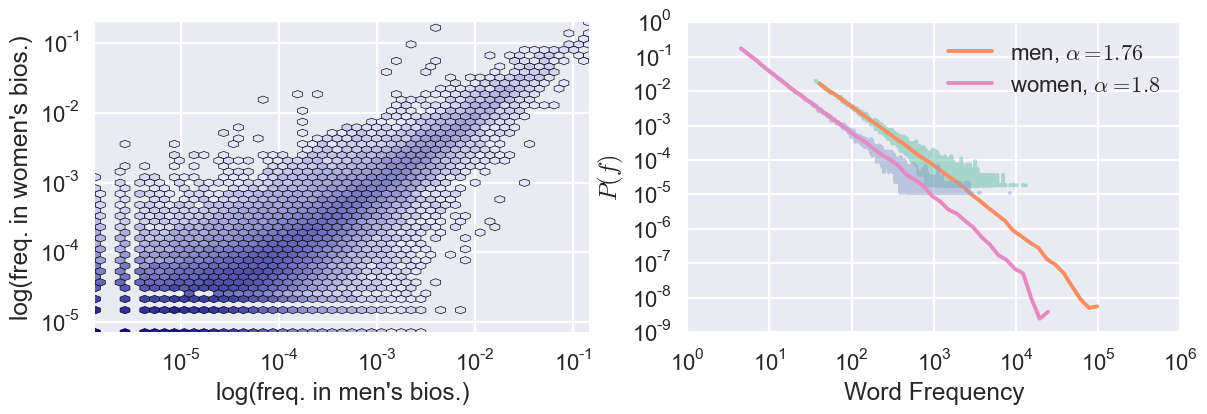}
\caption{A density hexbin plot of word frequencies in men/women's biographies (left), and the PDF of word frequency distribution according to gender (right).}
\label{fig:word-frequencies}
\end{figure}
}}

\begin{\widefigure}[tbp]
\centering
\begin{subfigure}[b]{0.49\linewidth}
        \includegraphics[width=\linewidth]{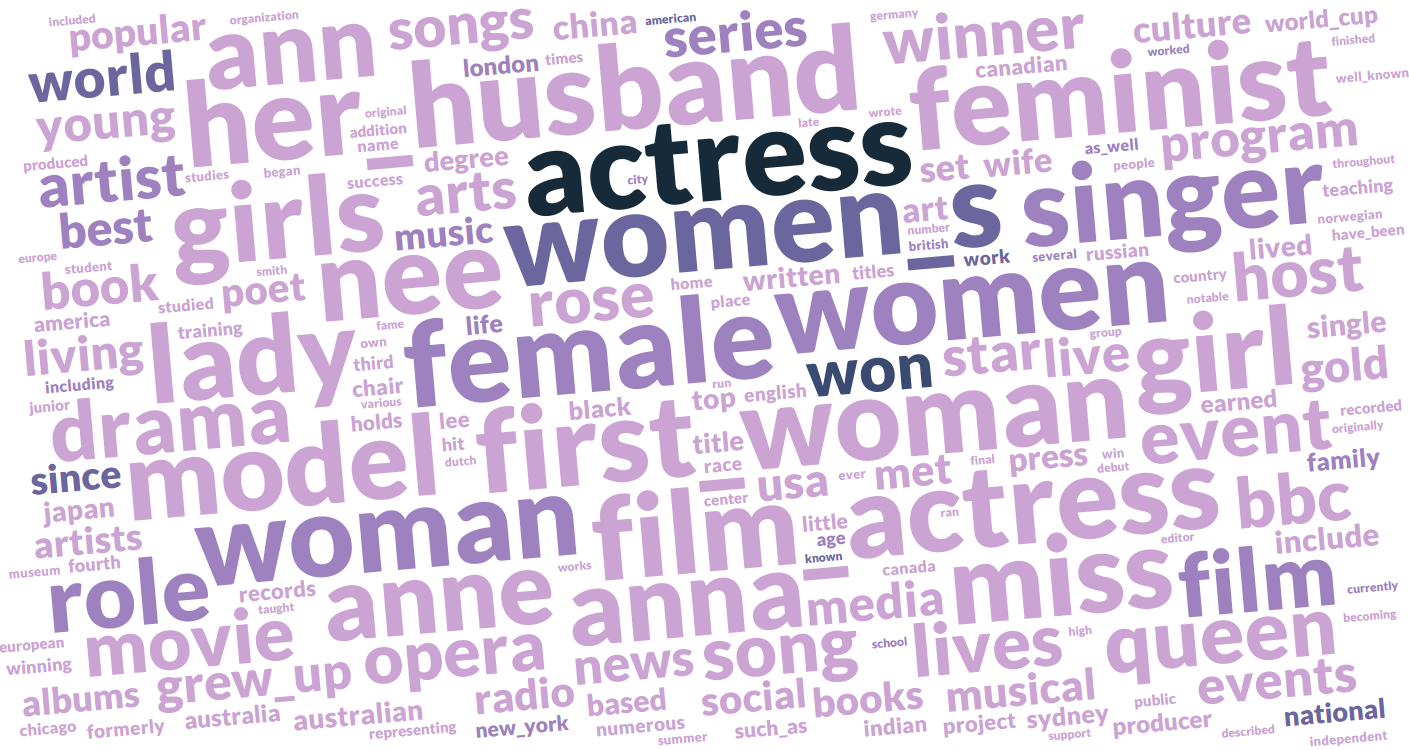}
\end{subfigure}%
\begin{subfigure}[b]{0.49\linewidth}
        \includegraphics[width=\linewidth]{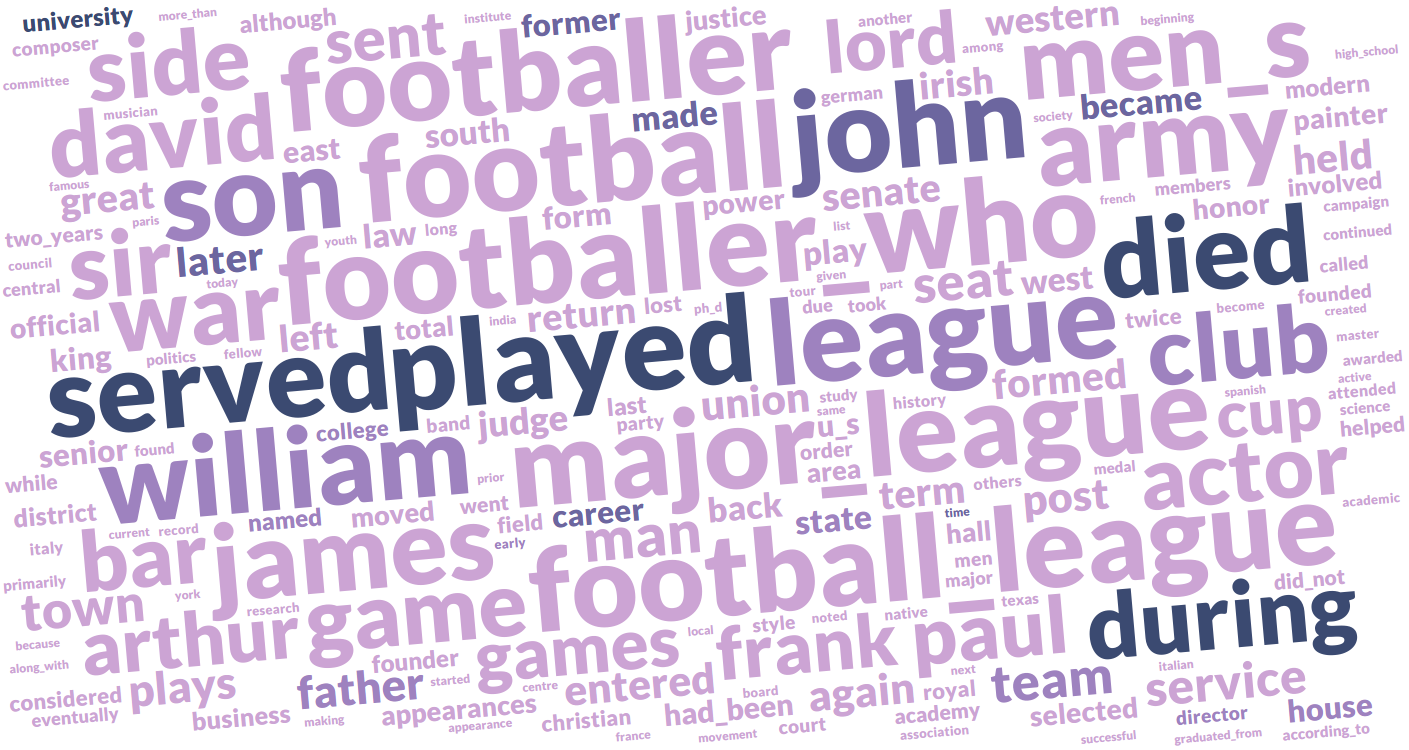}
\end{subfigure}%
\caption{Words most associated with women (left) and men (right), estimated with \textit{Pointwise Mutual Information}. Font size is inversely proportional to PMI rank. Color encodes frequency (the darker, the more frequent).}
\label{fig:wordclouds}
\end{\widefigure}

In this section we explore the characterization of women and men from a lexical perspective.
We analyze the vocabulary used in the overview of each biography through word frequency, and we use the estimated frequencies to find which words are associated with each gender.
To estimate relative frequencies, words were considered once per biography, and we estimated bi-gram word collocations to identify composite concepts (\eg, \textit{New York}).
We obtained a vocabulary of size $V_{m} = $ 1,013,305 for men, $V_{w} = $ 376,737 for women, with $V = $ 272,006 common words. 

Figure~\ref{fig:word-frequencies} displays a density plot of word frequency, and the Probability Density Functions (PDFs) for both genders. The frequency distributions are similar across genders. 
Word frequencies in the common vocabulary for both genders follow a Zipf distribution $P(f) \sim f^{-\alpha}$ with similar exponents $\alpha \approx 1.8$, consistent with the value found by \citet{serrano2009modeling}.
In addition, frequency \wrt gender presents a high rank-correlation $\rho = 0.65$ (p $< 0.001$). For reference, consider that the inter-language rank correlation of words with the same meaning across languages is 0.54 \cite{calude2011we}. This implies that words share meanings when referring to men and women.

{\ifthenelse{\boolean{inthesis}}{%
\subsection{Associativity of Words with Gender}
}{%
\spara{Associativity of Words with Gender}
}}%
To explore which words are more strongly associated with each gender, we measure \textit{Pointwise Mutual Information} \cite{church1990word} over the set of vocabulary in both genders. PMI is defined as:
\[
\mbox{PMI}(c, w) = \log \frac{p(c, w)}{p(c) p(w)}
\]
where $c$ is a class (\textit{men} or \textit{women}), and $w$ is a word. 
The probabilities can be estimated from the proportions of biographies about men and women, and the corresponding proportions of words and bi-grams.
Since PMI overweights words with very small frequencies, we consider only words that appear in at least 1\% of men or women biographies.

Associativity results are displayed as word clouds in Figure~\ref{fig:wordclouds}.
The top-15 words associated to each gender are (relative frequency in parentheses): 

\begin{itemize}
 \item Women: \textit{actress} (15.9\%), \textit{women's} (8.8\%), \textit{female} (5.6\%), \textit{her husband} (4.1\%), \textit{women} (5.3\%), \textit{first woman} (1.9\%), \textit{film actress} (1.6\%), \textit{her mother} (1.8\%), \textit{woman} (4.4\%), \textit{nee} (3.6\%), \textit{feminist} (1\%), \textit{miss} (1.9\%), \textit{model} (3.3\%), \textit{girls} (1.5\%) and \textit{singer} (6.5\%).
 \item Men: \textit{played} (14.2\%), \textit{footballer who} (3.0\%), \textit{football} (4.5\%), \textit{league} (5.9\%), \textit{john} (7.9\%), \textit{major league} (1.8\%), \textit{football league} (1.6\%), \textit{college football} (1.5\%), \textit{son} (7\%), \textit{football player} (2.2\%), \textit{footballer} (2\%), \textit{served} (11.7\%), \textit{william} (4.6\%), \textit{national football} (2\%) and \textit{professional footballer} (1\%).
\end{itemize}

Clearly, the words most associated with men are related to sports, football in particular, which refers to both popular sports of soccer and American football (recall from Table \ref{table:n-biographies-all} that \textit{Athlete} is the largest subclass of \textit{Person}).
For women, the most associated words are related to arts (recall from Table~\ref{table:n-biographies-all} that \textit{Artist} is the second largest subclass of \textit{Person}), gender (\textit{women's}, \textit{female}, \textit{first woman}, \textit{feminist}), and family roles (\textit{her husband}, \textit{her mother}, \textit{nee}\footnote{Adjective used when giving a former name of a woman.}).
This is consistent with the results from the meta-data analysis, where women are more likely to have a \textit{spouse} attribute in their infoboxes (see Table \ref{table:biography-meta-data}), and with the results of \citet{bamman2014unsupervised}.

{\ifthenelse{\boolean{inthesis}}{%
\subsection{Gender Differences in Semantic Categories of Words}
}{%
\spara{Gender Differences in Semantic Categories of Words}
}}%
Words most associated to each gender might belong to categories that are hard to compare, given their richness and complexity.
We use the \textit{Linguistic Inquiry and Word Count} dictionary of semantic categories to find if different genders have different characterizations according to those semantic categories.
The LIWC dictionary includes, for each category (and its corresponding subcategories), a list of words and prefixes that match relevant words.
We consider the following pertinent categories to our context: \textit{Social Processes}, \textit{Cognitive Processes}, \textit{Biological Processes}, \textit{Work Concerns} and \textit{Achievement Concerns}.
To generate the final dictionaries from the vocabulary, we matched the prefixes in our corpus and performed manual cleaning of noisy keywords like place names (\eg, \textit{Virginia} matches \textit{virgin*} from the \textit{sexual} category), surnames (\textit{Lynch} matches \textit{lynch*} from the \textit{death} category), and words with unrelated meanings.
In total, our cleaned dictionary contained 2,877 words.

\begin{\bigtable}[tb]
\scriptsize
\caption{Word frequency in biography overviews. For each LIWC category we report vocabulary size, median frequencies, the result of a Mann-Whitney \textit{U} test, and the three most frequent words. \textit{M} and \textit{W} mean men and women, respectively.}
\begin{tabulary}{\linewidth}{LRCCCLL}
\toprule
   Category &     V & Median (M) & Median (W) &        U &                                             Top-3 (M) &                                            Top-3 (W) \\
\midrule
     social &   498 &     0.04\% &     0.05\% &    -1.12 &           team (7.5\%), son (7.0\%), received (5.1\%) &     daughter (6.8\%), received (5.9\%), role (5.8\%) \\
  -- family &    43 &     0.03\% &     0.09\% &    -0.85 &           son (7.0\%), father (5.0\%), family (3.9\%) &     daughter (6.8\%), family (4.7\%), father (3.7\%) \\
  -- friend &    33 &     0.05\% &     0.05\% &    -0.58 &       fellow (2.0\%), friend (0.8\%), partner (0.8\%) &      fellow (1.8\%), partner (1.1\%), friend (0.8\%) \\
  -- humans &    59 &     0.13\% &     0.17\% &    -1.34 &         people (2.4\%), man (2.2\%), children (1.8\%) &      female (5.6\%), women (5.3\%), children (4.5\%) \\
    cogmech &  1045 &     0.02\% &     0.02\% &    2.04* &          became (10.8\%), known (9.8\%), made (8.1\%) &        known (10.3\%), became (9.2\%), since (8.1\%) \\
 -- insight &   354 &     0.02\% &     0.02\% &     0.73 &        became (10.8\%), known (9.8\%), become (2.2\%) &       known (10.3\%), became (9.2\%), become (2.0\%) \\
   -- cause &   182 &     0.02\% &     0.02\% &     1.31 &            made (8.1\%), since (6.2\%), based (3.3\%) &           since (8.1\%), made (6.7\%), based (4.2\%) \\
 -- discrep &    57 &     0.02\% &     0.02\% &     0.06 &  outstanding (0.4\%), wanted (0.3\%), besides (0.3\%) &    outstanding (0.5\%), wanted (0.4\%), hope (0.4\%) \\
  -- tentat &   151 &     0.01\% &     0.01\% &     0.85 &      appeared (3.2\%), mainly (0.9\%), mostly (0.8\%) &  appeared (6.8\%), appearing (1.1\%), mainly (0.8\%) \\
 -- certain &   110 &     0.03\% &     0.02\% &     0.92 &         law (2.7\%), total (1.1\%), completed (1.0\%) &         law (2.0\%), ever (1.0\%), completed (0.9\%) \\
   -- inhib &   229 &     0.01\% &     0.01\% &     1.75 &   held (4.2\%), conservative (0.7\%), control (0.5\%) &         held (3.1\%), hold (0.6\%), opposite (0.5\%) \\
    -- incl &     7 &     0.25\% &     0.29\% &    -0.06 &         addition (1.5\%), open (0.8\%), close (0.6\%) &        addition (1.7\%), open (1.0\%), close (0.4\%) \\
    -- excl &     6 &     0.11\% &     0.07\% &     0.48 &           except (0.2\%), whether (0.2\%), vs (0.1\%) &          except (0.2\%), whether (0.1\%), vs (0.1\%) \\
        bio &   638 &     0.01\% &     0.01\% &    -1.63 &            life (3.9\%), head (2.5\%), living (1.1\%) &           life (4.7\%), love (1.9\%), living (1.7\%) \\
    -- body &   193 &     0.01\% &     0.01\% &    -0.60 &              head (2.5\%), body (0.6\%), face (0.5\%) &             head (1.5\%), body (0.8\%), face (0.6\%) \\
  -- health &   274 &     0.01\% &     0.01\% &    -0.40 &        life (3.9\%), living (1.1\%), hospital (0.9\%) &         life (4.7\%), living (1.7\%), health (1.2\%) \\
  -- sexual &   105 &     0.00\% &     0.01\% &  -3.02** &            love (0.8\%), passion (0.2\%), gay (0.2\%) &           love (1.9\%), sex (0.5\%), lesbian (0.3\%) \\
  -- ingest &   122 &     0.01\% &     0.01\% &    -0.51 &             water (0.4\%), food (0.3\%), cook (0.2\%) &            food (0.5\%), water (0.4\%), cook (0.3\%) \\
       work &   570 &     0.04\% &     0.03\% &     1.12 &          career (9.5\%), team (7.5\%), worked (6.4\%) &       career (8.1\%), worked (6.6\%), school (6.1\%) \\
    achieve &   364 &     0.05\% &     0.04\% &     1.06 &             won (8.7\%), team (7.5\%), worked (6.4\%) &           won (13.0\%), worked (6.6\%), team (5.5\%) \\
\bottomrule
\end{tabulary}

\label{table:liwc-frequencies}
\end{\bigtable}

\begin{\bigtable}[tb]
\scriptsize
\caption{Word burstiness in full biographies for LIWC categories. Columns are analog to Table \ref{table:liwc-frequencies}.}
\begin{tabulary}{\linewidth}{LRCCCLL}
\toprule
   Category &     V &  Median (M) &  Median (W) &       U &                                             Top-3 (M) &                                            Top-3 (W) \\
\midrule
     social &   498 &        1.21 &        1.22 &    0.21 &                 band (3.63), team (3.38), game (2.95) &               team (3.40), women (3.14), role (2.96) \\
  -- family &    43 &        1.31 &        1.35 &   -1.12 &              family (1.85), father (1.75), son (1.64) &          family (2.02), mother (1.98), granny (1.87) \\
  -- friend &    33 &        1.23 &        1.26 &   -1.06 &           friendly (1.86), buddy (1.66), guest (1.59) &        guest (1.75), fellowship (1.54), buddy (1.53) \\
  -- humans &    59 &        1.35 &        1.44 &   -1.00 &                  sir (2.39), human (2.02), man (2.02) &                women (3.14), mrs (2.33), lady (2.25) \\
    cogmech &  1045 &        1.12 &        1.12 &  2.85** &              open (2.37), law (2.36), decision (2.34) &            open (3.28), revelator (2.75), law (2.31) \\
 -- insight &   354 &        1.13 &        1.12 &    1.75 &          decision (2.34), logic (2.15), became (1.88) &       revelator (2.75), became (1.86), ponder (1.86) \\
   -- cause &   182 &        1.15 &        1.13 &   2.17* &          force (2.05), made (1.92), production (1.85) &  causation (2.29), outcome (2.04), production (1.82) \\
 -- discrep &    57 &        1.10 &        1.14 &   -1.05 &     desir (1.49), outstanding (1.48), idealism (1.45) &      outstanding (2.03), wanna (1.58), oughta (1.55) \\
  -- tentat &   151 &        1.12 &        1.10 &    1.86 &      mysterium (1.96), puzzle (1.70), appeared (1.68) &          appeared (2.02), bet (1.67), overall (1.63) \\
 -- certain &   110 &        1.11 &        1.10 &    1.62 &                law (2.36), total (2.14), truth (1.50) &             law (2.31), reality (1.55), total (1.52) \\
   -- inhib &   229 &        1.10 &        1.10 &    1.09 &     fencing (2.28), security (1.92), defensive (1.89) &          fencing (2.20), safe (2.16), blocker (2.02) \\
    -- incl &     7 &        1.27 &        1.29 &   -0.45 &              open (2.37), inside (1.30), close (1.30) &           open (3.28), close (1.31), additive (1.30) \\
    -- excl &     6 &        1.27 &        1.20 &    0.48 &              vs (2.17), versus (1.32), whether (1.31) &             vs (1.75), versus (1.35), whether (1.24) \\
        bio &   638 &        1.26 &        1.25 &    1.87 &    choke (5.18), lymphology (3.50), pelvimeter (3.00) &            love (2.52), prostatic (2.50), hiv (2.33) \\
    -- body &   193 &        1.27 &        1.26 &    1.24 &             pelvimeter (3.00), hip (2.15), pee (2.03) &     prostatic (2.50), pelvimeter (2.00), tits (1.98) \\
  -- health &   274 &        1.24 &        1.24 &    1.33 &  choke (5.18), lymphology (3.50), chiropractic (2.92) &             hiv (2.33), choke (2.32), insulin (2.16) \\
  -- sexual &   105 &        1.27 &        1.31 &   -0.51 &                   gay (2.56), hiv (2.35), love (2.12) &            love (2.52), prostatic (2.50), hiv (2.33) \\
  -- ingest &   122 &        1.29 &        1.24 &    1.30 &                 chew (2.39), cook (2.22), coke (2.11) &          cookery (2.18), cooking (1.98), food (1.95) \\
       work &   570 &        1.23 &        1.20 &  2.62** &                gre (4.54), team (3.38), dotcom (2.98) &               pce (18.67), team (3.40), award (3.40) \\
    achieve &   364 &        1.15 &        1.15 &    0.54 &                  team (3.38), win (3.19), king (2.64) &               team (3.40), award (3.40), best (2.72) \\
\bottomrule
\end{tabulary}

\label{table:liwc-burstiness}
\end{\bigtable}

To compare the distribution of words in the semantic categories, we employed two metrics: relative frequency in overviews, as previously done with PMI, and burstiness in the full text. 
Word frequencies identify how language is used differently to characterize men and women in terms of semantic categories.
However, word frequency alone does not give insights on how those semantic categories portray a given biography, or in other words, the importance that editors give to those categories.
Burstiness is a measure of word importance in a single document according to the number of times it appears within the document, under the assumption that important words appear more than once (they appear in \textit{bursts}) when they are relevant in a given document.
We use the definition of burstiness from \citet{church1995poisson}:
\[
B(w) = \frac{E_{w}(f)}{P_{w}(f \geq 1)}
\]
where $E_{w}(f)$ is the mean number of occurrences of a given word $w$ per document, and $P_{w}(f \geq 1)$ is the probability that $w$ appears at least once in a document.
The differences in frequency and burstiness are tested using the Mann-Whitney \textit{U} test, which indicates if one population tends to have larger values than another. It is non-parametric, \ie, it does not assume normality.

\spara{Differences in Frequency}
Table \ref{table:liwc-frequencies} shows statistics related to word frequency in biography overviews for the LIWC categories. 
Note that, although the medians are very similar for each category, the \textit{U} test compares differences in the distribution instead of differences in means or medians. 
If the test revealed significant differences, we calculated the \textit{common language effect size} (ES) as the percentage of words that had a greater relative frequency for the dominant gender. 
Of the 20 categories under consideration, two of them (one top-level) shown significant differences between genders: \textit{cogmech} (\textit{cognitive processes}, ES = 63\%) is dominated by men, while \textit{sexual} (\textit{sexual processes}, subcategory of \textit{biological processes}, ES = 85\%) is dominated by women.

\spara{Differences in Burstiness}
Burstiness distributions in full biographies per semantic category are displayed in Table \ref{table:liwc-burstiness}. 
There are three (two top-level) categories with significant differences, both dominated by men:  
\textit{cogmech} (\textit{cognitive processes}, ES = 60\%), its subcategory \textit{cause} (\textit{causal processes}, ES = 71\%), and \textit{work} (\textit{work concerns}, ES = 64\%).

\spara{Overview of Results}
In summary, in this section we found that words have similar meaning when referring to both genders, that there are qualitative differences in words most associated to them, and that a small number of the semantic categories show significant differences.
Although this implies more similarities than differences in characterization of women and men, in the discussion section we elaborate over the importance of such differences and the implications of these findings.

{\ifthenelse{\boolean{inthesis}}{\afterpage{\clearpage}}{}

\section{Network Properties}

To study structural properties of biographies, we first built a directed network of biographies from the links between articles in the \textit{Person} DBPedia class.
This empirical network was compared with several null graphs that, by construction, preserve different known properties of the original network. 
This allows us to attribute observed structural differences between genders either to empirical fluctuations in such properties, such as the heterogeneous importance of historical figures, or to gender bias.
To do so, we consider PageRank, a measure of node centrality based on network connectivity \cite{brin1998anatomy,Fortunato08internetmath}. 

{\ifthenelse{\boolean{inthesis}}{%
\subsection{Empirical Network and Null Models}
}{%
\spara{Empirical Network and Null Models}
}}%
We study the properties of the directed network constructed from the links between 893,380 biographical articles in the \textit{Person} class.
After removing 192,674 singleton nodes, the resulting graph has 700,706 nodes and 4,153,978 edges. We use this graph to construct the following null models:

\begin{itemize}
 \item \textit{Random}. 
We shuffle the edges in the original network. 
For each edge (u,v), we select two random nodes (i,j) and replace (u,v) by (i,j). The resulting network is a random graph with neither the heterogeneous degree distribution nor the clustered structure that the Wikipedia graph is known to have \cite{zlatic2006wikipedias}.
 \item \textit{In-Degree Sequence}. 
We generate a graph that preserves the in-degree sequence (and therefore the heterogeneous in-degree distribution) of the original network by shuffling the sources of the edges. 
For each edge (u,v), we select a random node (i) and rewire (u,v) to (i,v). Each node has the same in-degree, or popularity, as the corresponding biography. 
 \item \textit{Out-Degree Sequence}. 
We generate a graph that preserves the out-degree sequence (and therefore the out-degree distribution) of the original network by shuffling the targets of the edges. 
For each edge (u,v) select a random node (j) and rewire (u,v) to (u,j).
 \item \textit{Full Degree Sequence}. 
We generate a graph that preserves both in-degree and out-degree sequences (and therefore both distributions) by shuffling the structure in the original network. 
For a random pair of edges ((u,v), (i,j)) rewire to ((u,j), (i,v)). 
We repeat this shuffling as many times as there are edges. Note that although the in- and out-degree of each node is unchanged, the degree correlations and the clustering are lost.
 \item \textit{Small World}. 
 We generate a undirected small world graph using the model by \citet{watts1998collective}. 
This model interpolates a random graph and a lattice in a way that preserves two  properties of small world networks: average path length and clustering coefficient.
\end{itemize}

All null models have the same number of nodes $n = $ 700,706 and approximately the same mean degree $k \approx 4$ as the empirical network.
{\ifthenelse{\boolean{inthesis}}{The Small World model has a parameter $\beta = 0.34$ representing the probability of rewiring each edge. Its value was set using the Brent root finding method in such a way as to recover the clustering coefficient of the original network.}{}}

{\ifthenelse{\boolean{inthesis}}{
\begin{table*}[tb]
\scriptsize
\caption{Comparison of edge proportions between genders in the empirical biography network and the null models. \textit{M} and \textit{W} mean men and women, respectively. All models share the same number of nodes, $n =$ 700,706. }
\centering
\begin{tabulary}{\linewidth}{lRRRRLRRLR}
\toprule
{} &  Clust. Coef. &    Edges & M to M & M to W & $\chi^{2}$ (M to W) & W to M & W to W & $\chi^{2}$ (W to W) &  SFR \\
\midrule
Observed       &           0.16 &  4,106,916 &        90.05\% &         9.95\% &                2.38 &        62.19\% &        37.81\% &            37.83*** & 6.55 \\
Small World    &           0.16 &  2,775,372 &        84.45\% &        15.55\% &                0.00 &        84.15\% &        15.85\% &                0.01 & 5.41 \\
Random         &           0 &  4,106,916 &        84.41\% &        15.59\% &                0.00 &        84.39\% &        15.61\% &                0.00 & 5.41 \\
In Deg. Seq.   &           0 &  4,106,916 &        85.36\% &        14.64\% &                0.06 &        85.27\% &        14.73\% &                0.05 & 5.75 \\
Out Deg. Seq.  &           0 &  4,106,916 &        84.43\% &        15.57\% &                0.00 &        84.37\% &        15.63\% &                0.00 & 5.42 \\
Full Deg. Seq. &           0 &  4,106,916 &        85.34\% &        14.66\% &                0.06 &        85.39\% &        14.61\% &                0.06 & 5.74 \\
\bottomrule
\end{tabulary}

\label{table:biography-network-properties}
\end{table*}
}{
\begin{table*}[tb]
\scriptsize
\caption{Comparison of the empirical biography network and the null models. \textit{M} and \textit{W} mean men and women, respectively.}
\centering
\begin{tabulary}{\linewidth}{lRRRRRLRRLR}
\toprule
{} &   Nodes &    Edges &  Clust. Coeff. & Edges (M to M) & Edges (M to W) & $\chi^{2}$ (M to W) & Edges (W to M) & Edges (W to W) & $\chi^{2}$ (W to W) &  SFR \\
\midrule
Observed       &  693843 &  4106916 &           0.16 &        90.05\% &         9.95\% &                2.38 &        62.19\% &        37.81\% &            37.83*** & 6.55 \\
Small World    &  693843 &  2775372 &           0.16 &        84.45\% &        15.55\% &                0.00 &        84.15\% &        15.85\% &                0.01 & 5.41 \\
Random         &  693843 &  4106916 &           0.00 &        84.41\% &        15.59\% &                0.00 &        84.39\% &        15.61\% &                0.00 & 5.41 \\
In Deg. Seq.   &  693843 &  4106916 &           0.00 &        85.36\% &        14.64\% &                0.06 &        85.27\% &        14.73\% &                0.05 & 5.75 \\
Out Deg. Seq.  &  693843 &  4106916 &           0.00 &        84.43\% &        15.57\% &                0.00 &        84.37\% &        15.63\% &                0.00 & 5.42 \\
Full Deg. Seq. &  693843 &  4106916 &           0.00 &        85.34\% &        14.66\% &                0.06 &        85.39\% &        14.61\% &                0.06 & 5.74 \\
\bottomrule
\end{tabulary}

\label{table:biography-network-properties}
\end{table*}
}}

\spara{Gender, Link Proportions and Self-Focus Ratio}
For each graph, we estimated the proportion of links from gender to gender, and we tested those proportions against the expected proportions of men and women present in the dataset using a chi-square test.
Table \ref{table:biography-network-properties} shows the results. 
None of the null models show any bias in link proportions. 
The observed graph, on the other hand, shows a significant difference in the proportion of links from women biographies.  
In particular, articles about women tend to link to other women biographies more than expected ($\chi^2 = 40.54, p < 0.001$, Cohen's $w = 0.76$). 
Men biographies show a greater proportion of links to men and a lesser proportion to women than expected, but the difference is not statistically significant, although it has an impact on the estimated \textit{Self-Focus Ratio} \cite{hecht2009measuring}.
In our context, this ratio is defined as the relation between the sum of PageRank for men and the sum of PageRank for women. 
A SFR above 1 confirms the presence of self-focus, which, given the proportions of men and women in the dataset, is expected. In fact, given those proportions, the expected SFR is 5.41. 
Note that the null models have similar SFRs to the expected value, in contrast with the observed model with SFR of 6.55.

{\ifthenelse{\boolean{inthesis}}{%
\subsection{Biography Importance}
}{%
\spara{Biography Importance}
}}%
As an approximation for historical importance in our biography network we considered the ranking of biographies based on their PageRank values.
{\ifthenelse{\boolean{inthesis}}{%

\begin{figure}[p]
\centering
\includegraphics[width=\linewidth]{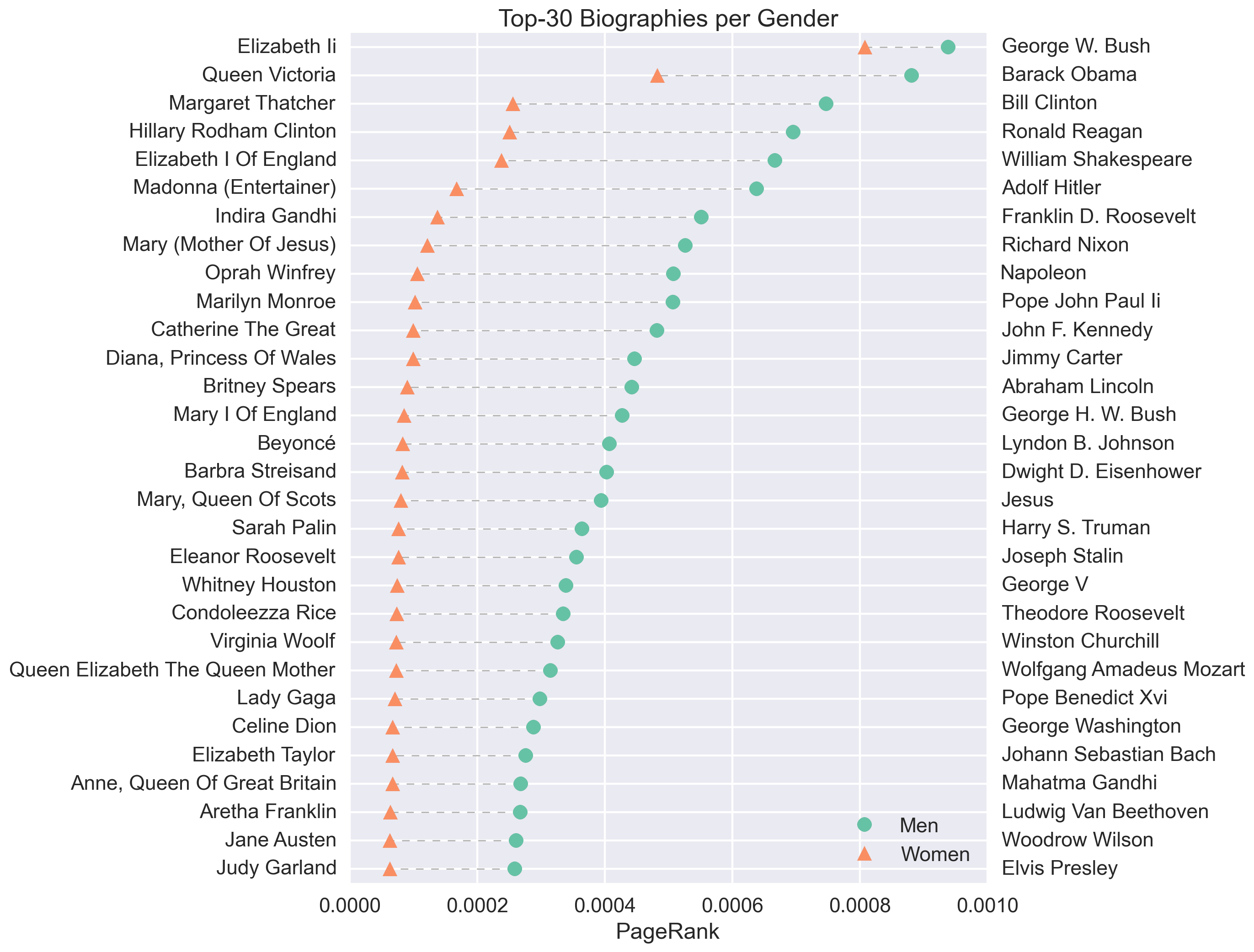}
\caption{Top-30 biographies per gender according to PageRank. }
\label{fig:gender_pagerank}
\end{figure}

\begin{figure}[p]
\centering
\includegraphics[width=\linewidth]{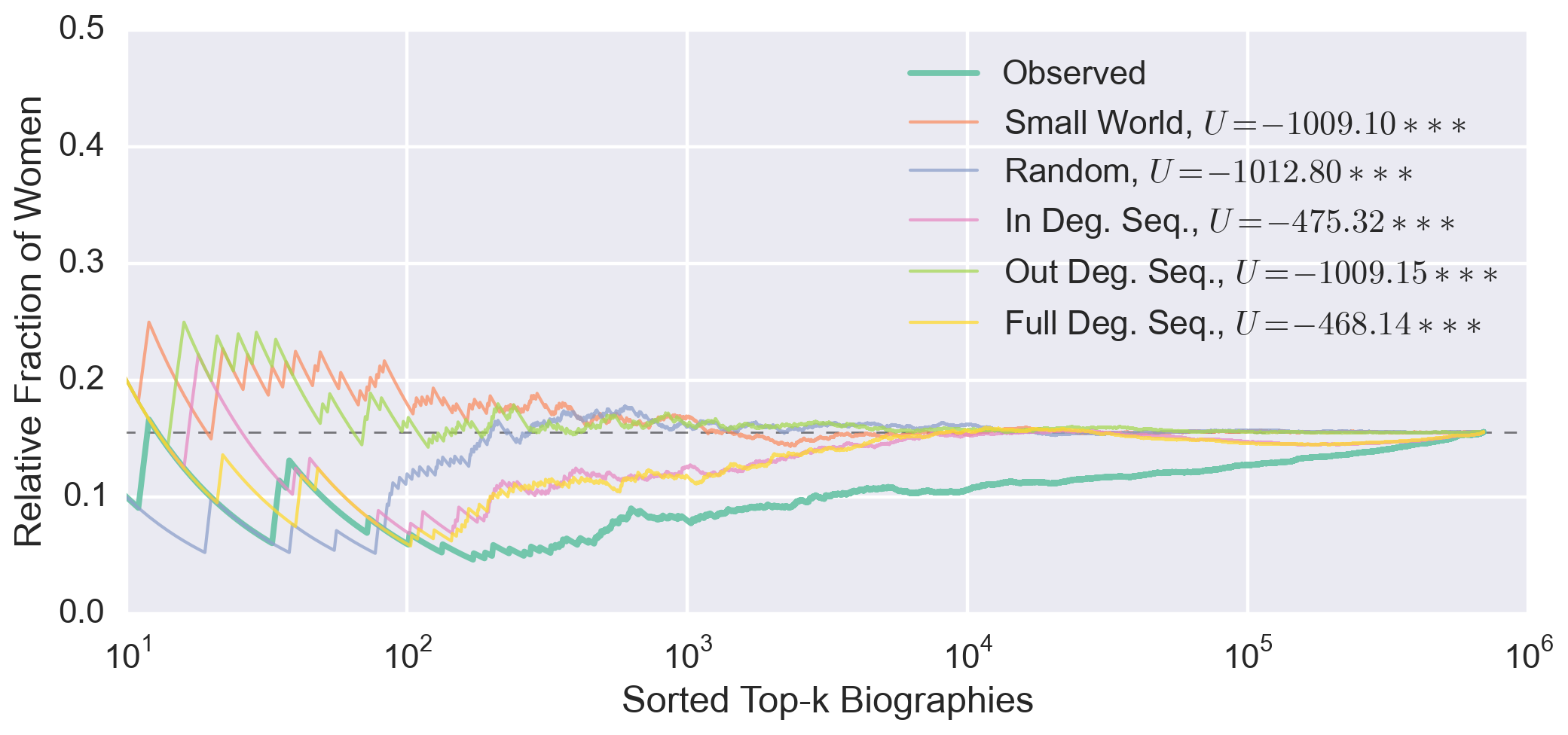}
\caption{Women fraction in top biographies sorted by PageRank.}
\label{fig:ccdf-pagerank}
\end{figure}

Figure \ref{fig:gender_pagerank} displays the top-30 men and women according to their PageRank. 
Although the highest score entities present comparable scores, women present a faster decay than men. 
For instance, \textit{Pope John Paul II}\footnote{\url{https://en.wikipedia.org/wiki/Pope_John_Paul_II}} (\#10) has higher score than \textit{Queen Victoria}\footnote{\url{https://en.wikipedia.org/wiki/Queen_Victoria}} (\#2), and \textit{Elvis Presley}\footnote{\url{https://en.wikipedia.org/wiki/Elvis_Presley}} (\#30) has higher score than \textit{Hillary Rodham Clinton}\footnote{\url{https://en.wikipedia.org/wiki/Hillary_Rodham_Clinton}} (\#4).
Our results are coherent with previous work: \citet{aragon2012biographical} is more similar to ours because they consider PageRank only, while \citet{whosbigger} considers other additional factors when ranking. 

}{%

\begin{figure}[tb]
\centering
\includegraphics[width=\linewidth]{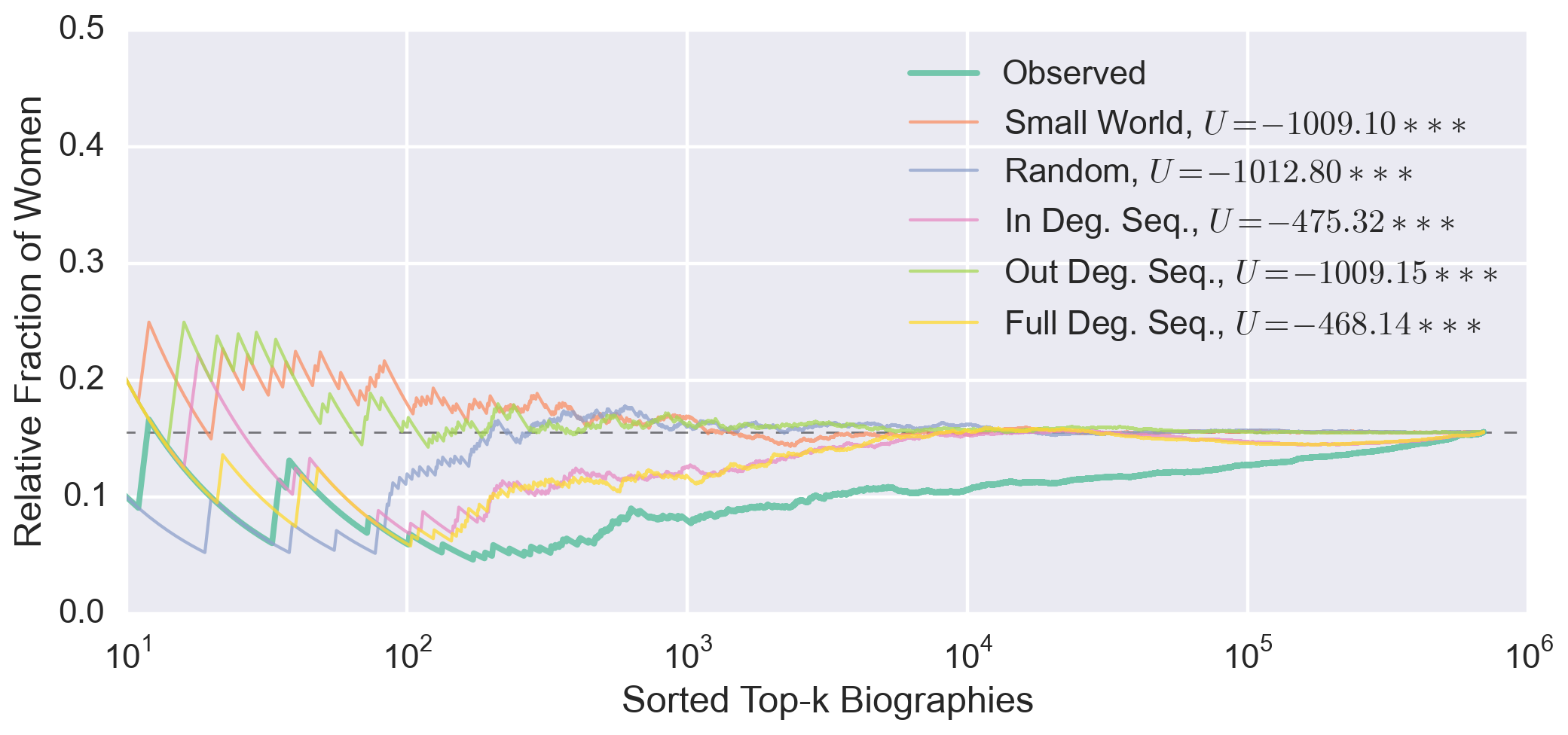}
\caption{Women fraction in top biographies sorted by PageRank.}
\label{fig:ccdf-pagerank}
\end{figure}

}}%
To compare the observed distribution of PageRank by gender to those of the null models,  
we analyzed the fraction of women biographies among the top-$r$ articles by PageRank, for $r \in [10,~700,706]$ (\ie, we considered only nodes with edges). 
In the absence of any kinds of bias, whether endogenous to Wikipedia or exogenous, one would expect the fraction of women to be around 15\% (the overall proportion of women biographies) irrespective of $r$. In the presence of correlations between popularity or historical importance and gender, we expect the ratio to fluctuate. But such fluctuations would also be observed in the null models. 

The results are shown in Figure \ref{fig:ccdf-pagerank}.
While the null models stabilize around the expected value by $r \leq 10^4$, the proportion of women in the observed network reaches 15\% only when the entire dataset is considered. This systematic under-representation of women among central biographies is not mirrored in the null models. 
We tested the differences between observed and null models using a Mann-Whitney \textit{U} test, and found that the observed model is always significantly different (\textit{U} values shown in Figure \ref{fig:ccdf-pagerank}, p $< 0.001$ for all pairwise comparisons with the observed model, Holm-Sidak corrected).
This implies a biased behavior that cannot be explained by any of the heterogeneities in the structure of the network preserved by the null models.
For instance, even if men biographies tended to have more incoming links (as they do), or to be more densely clustered, those factors would not explain the lower centrality observed in women biographies. 

{\ifthenelse{\boolean{inthesis}}{\afterpage{\clearpage}}{}

\section{Discussion}
Even though we found more similarities than differences in characterization, in this section we contextualize those differences in social theory and history. We do this to understand why such differences exist, and whether they can be attributed to bias in Wikipedia or to a reflection of western society.

\spara{Meta-data}
We found that there are statistically significant differences in biographies of men and women. 
Most of them can be explained because of the different areas to which men and women belong (mostly \textit{sports} and \textit{arts}, respectively), as well as the recency of women profiles available on Wikipedia.
Other differences, like article length and article out-degree, although significant, have very small effect sizes, and depend on the person class being analyzed.
 
The greater frequency of the \textit{spouse} attribute in women can be interpreted as specific gender roles attributed to women.
A similar result on \textit{Implicit Association} was obtained by \citet{nosek2002harvesting}, as they found that Internet visitors tended to associate women to family and arts.
Arguably, an alternative explanation is that people in the arts could be more likely to marry a notable spouse than people in sports. 
Yet, we found that the most common  class was the generic one not assigned to any of those categories.

In terms of time, we found that the year 1943 marked a hit on the growth of women presence.
According to \citet{strauss1991generations}, the post-war \textit{Baby Boomers} generation started in 1943. The following generations are \textit{Generation X} (1961--1981) and \textit{Millenials} (1982--2004). 
The social and cultural changes embraced by people from those generations, plus the increased availability of secondary sources, might explain this growth. 
The growth started in dates nearby the French Revolution (1789--1799), where women had an important role, although they were oppressed after it \cite{abray1975feminism}. During these years seminal works about feminist philosophy and women's rights were published, like 
the works of \textit{Mary Wollstonecraft} (1792) and 
\textit{Olympe de Gouges} (1791). 
It is reasonable to assume that these historical events paved the way for women to become more notable.

\spara{Language}
We found that the words most associated with men are mostly about sports, while the words most associated with women are to arts, gender and family. 
Of particular interest are two concepts strongly associated with women: \textit{her husband} and \textit{first woman}.
These results are arguably indicative of systemic bias: the usage of \textit{her husband} was found in concordance with our meta-data results and previous work by \citet{bamman2014unsupervised}, and the already mentioned work on \textit{Implicit Association} \cite{nosek2002harvesting}.
These results can be contextualized in terms of \textit{stereotyping theory} \cite{pratto2007communication}, as they categorize women, either as norm breaking (being the first is an exception to the norm) or as with predefined roles (being wives). 
As \citet{fiske1990continuum} indicate in their \textit{continuum model of impression formation}, such categorization makes individuals more prone to stereotyping than those who are not categorized.
The usage of \textit{first woman} might indicate notability, but it also has been seen as an indicator of gender bias, as indicated by the Bechdel-inspired \textit{Finkbeiner-test}\footnote{\url{http://www.doublexscience.org/the-finkbeiner-test/}} about scientific women, where it is explicitly mentioned that an article about a woman does not pass the test if it mentions \textit{``How she's the `first woman to \ldots'''}
Despite being informal, the Finkbeiner-test raises awareness on how gender becomes more important than the actual achievements of a person.

To formalize the PMI analysis, we performed analysis based on semantic dictionaries of words.
According to \citet{nussbaum1995objectification}, one possible indicator of \textit{objectification} is the \textit{``denial of subjectivity: the objectifier treats the object as something whose experience and feelings (if any) need not be taken into account.''}
This idea is supported as, in the overviews, men are more frequently described with words related to their \textit{cognitive processes}, while women are more frequently described with words related to \textit{sexuality}. In the full biography text, the \textit{cognitive processes} and \textit{work concerns} categories are more bursty in men biographies, meaning that those aspects of men's lives are more important than others at the individual level. 

\spara{Presence and Centrality of Women}
Women biographies tend to link more to other women than to men, a disproportion that might be related with women editing women biographies in Wikipedia, one of the reported interests of women editors \cite{wikisurvey}.
Since we are considering notable people, it is known that men and women's networks evolve differently through their careers \cite{jacobs1989revolving}, not to mention the set of life-events that influence those changes like child-bearing and marriage (see a in-depth discussion by \citet{smith1993you}). Thus, link proportion between women cannot be attributed to bias in Wikipedia, as it seems to be more a reflection of what happens in the physical world.

We found that network structure is biased in a way that gives more importance to men than expected, by comparing the distribution of PageRank across genders.
The articles with highest centrality, or historical importance \cite{aragon2012biographical}, tend to be predominantly about men, beyond what one could expect from the structure of the network. 
As shown in Figure \ref{fig:ccdf-pagerank}, there are women biographies with high centrality, but their presence is not a sign of an unbiased network: 
\textit{``the successes of some few privileged women neither compensate for nor excuse the systematic degrading of the collective level; and the very fact that these successes are so rare and limited is proof of their unfavorable circumstances''} \cite{de2012second}.

\subsection{Implications}
At this point, considering the \textit{gender gap} that affects Wikipedia \citep{hill2013wikipedia}, it is pertinent to recall the concept of \textit{feminine mystique} by \citet{friedan2010feminine}, developed from the analysis of women's magazines from the 50s in the United States, which were edited by men only.
Fortunately, as discussed earlier, we have found women in different fields, mostly \textit{arts}, in contrast to the \textit{``Occupation: Housewife''} identified by \citet{friedan2010feminine}, as well as more similarities in characterization than differences.
Moreover, the presence of women is increasing steadily and most of the differences found are not from an inherent bias in Wikipedia.
Nevertheless, the identified language differences objectify women and the network structure diminishes their findability and centrality. Hence, the gender bias in Wikipedia is not just a matter of women participation in the community, because content and characterization of women is also affected.
This is important, for example, because Wikipedia is used as an educational tool \cite{konieczny2010teaching}, and \textit{``children learn which behaviors are appropriate to each sex by observing differences in the frequencies with which male and female models as groups perform various responses in given situations''} \cite{perry1979social}.

\spara{Editing Wikipedia and NPOV}
Critics may rightly say that by relying on secondary sources, Wikipedia just reflects the biases found in them.
However, editors are expected to write in their own words, \textit{``while substantially retaining the meaning of the source material''}\footnote{\url{https://en.wikipedia.org/wiki/Wikipedia:No_original_research}}, and thus, the differences found in terms of language that objectify women are chosen explicitly by them.
In this aspect, Wikipedia should provide tools that help editors to reduce sexism in language, for instance, by considering already existing manuals like \cite{apa_manual}.
Furthermore, their neutral point of view guidelines should be updated to explicitly include gender bias, because biased language is a clear violation of their guidelines. 

\spara{Affirmative Action for Women in Notability Guidelines}
The current notability guidelines for biographies in Wikipedia state: \textit{``1. The person has received a well-known and significant award or honor, or has been nominated for one several times. 2. The person has made a widely recognized contribution that is part of the enduring historical record in his or her specific field.''}\footnote{\url{https://en.wikipedia.org/wiki/Wikipedia:Notability_(people)\#Any_biography}}
However, the boundary between not being notable according to sources and exclusion from history is blurred when evaluating the notability of women.
For instance, consider a discussion about women in philosophy: \textit{``Feminist historians of philosophy have argued that the historical record is incomplete because it omits women philosophers, and it is biased because it devalues any women philosophers it forgot to omit. In addition, feminist philosophers have argued that the philosophical tradition is conceptually flawed because of the way that its fundamental norms like reason and objectivity are gendered male''} \cite{sep-feminism-femhist}.
Women, specially in historical contexts before 1943, should be targeted by affirmative actions that would allow them to appear in the content if they are not there, and be linked from other articles.
We acknowledge that this is not easy, because relaxing notability guidelines can open the door to original research, which is not allowed. 
However, a correctly defined affirmative strategy would allow to grow the proportion of women in Wikipedia, make women easier to find, both through search (as it increases relevance) and exploratory browsing.

\subsection{Summary, Limitations and Future Work}
We studied gender bias in Wikipedia biographies.
Our results indicate significant differences in meta-data, language, and network structure that can be attributed not only to the mirroring of the offline world, but also to gender bias endogenous to content generation in Wikipedia. 
Our contribution is a set of methodologies that detect and quantify gender bias with respect to content and structure, as well as a contextualization of the differences found in terms of feminist theory.
As concluding remark, we discussed that Wikipedia may wish to consider revising its guidelines, both to account for the non-findability of women and to encourage a less biased use of language, which is a violation of its neutral point of view guideline.

\spara{Limitations}
Our study has two main limitations. 
First, our focus is on the English Wikipedia, which is biased towards western cultures.
However, a parallel work to ours by \citet{wagner2015s} focused on hyperlingual quantitative analysis, and obtained similar results for other languages. Our methods can be applied in other contexts given the appropriate dictionaries with semantic categories, although our discussion remains to be applied, as it is culture-dependent.
The second limitation is a binary gendered view, but we believe this is a first step towards analyzing the gender dimension in content from a wider perspective, given the social theory discussion we have made.

\spara{Future Work}
At least three areas are ripe for further work. 
The first is the construction of editing tools for Wikipedia that would help editors  detect bias in content, and suggest appropriate actions. 
The second is a study of individual differences among contributors, as our work analyzed user generated content without considering \textit{who} published and edited it.
This aspect can be explored by analyzing how contributors discuss and edit content based on their gender and other individual factors. 
The last area is a further exploration of bias considering more fine-grained ontology classes and meta-data attributes. For instance, it may be possible that gender bias is stronger or weaker for different ontology classes (e.g., \textit{Scientist} vs. \textit{Artist}) or in biographies of people from different regions and religions. Finally it would be helpful to study whether gender bias depends on the quality of an article: does bias decrease with increasing number of edits or other measures of article maturity?

{
\spara{Acknowledgments}
We thank Daniela Alarcón for fruitful discussion and Luca Chiarandini for tool code.
This work was partially funded by Grant TIN2012-38741 (Understanding Social Media: An Integrated Data Mining Approach) of the Min. of Economy and Competitiveness of Spain.
}

\printbibliography

\end{document}